\documentclass[12pt, a4paper]{article}

\usepackage{amsmath}
\usepackage{amssymb}
\usepackage{amsfonts}
\usepackage{amsthm}
\usepackage{latexsym}
\usepackage{color}
\usepackage{epsfig}
\usepackage{hyperref}
\usepackage{longtable}
\usepackage{multirow}
\usepackage{lscape}

\newcommand{\te}{{T2E }}
\let\proglang=\textsf
\newcommand{\pkg}[1]{{\fontseries{b}\selectfont #1}}

\def\bi{\begin{itemize}}
\def\ei{\end{itemize}}

\begin{document}

\title{Treatment Effect Quantification for Time-to-event Endpoints -- Estimands, Analysis Strategies, and beyond}

\author{
    Kaspar Rufibach\footnote{Methods, Collaboration, and Outreach Group (MCO), Department of Biostatistics, Hoffmann-La Roche Ltd, Basel, Switzerland} \\
}

\date{\today}

\maketitle

\begin{abstract}
A draft addendum to ICH E9 has been released for public consultation in August 2017. The addendum focuses on two topics particularly relevant for randomized confirmatory clinical trials: estimands and sensitivity analyses. The need to amend ICH E9 grew out of the realization of a lack of alignment between the objectives of a clinical trial stated in the protocol and the accompanying quantification of the ``treatment effect'' reported in a regulatory submission.
We embed time-to-event endpoints in the estimand framework, and discuss how the four estimand attributes described in the addendum apply to time-to-event endpoints.
We point out that if the proportional hazards assumption is not met, the estimand targeted by the most prevalent methods used to analyze time-to-event endpoints, logrank test and Cox regression, depends on the censoring distribution.
We discuss for a large randomized clinical trial how the analyses for the primary and secondary endpoints as well as the sensitivity analyses actually performed in the trial can be seen in the context of the addendum. To the best of our knowledge, this is the first attempt to do so for a trial with a time-to-event endpoint. Questions that remain open with the addendum for time-to-event endpoints and beyond are formulated, and recommendations for planning of future trials are given. We hope that this will provide a contribution to developing a common framework based on the final version of the addendum that can be applied to design, protocols, statistical analysis plans, and clinical study reports in the future.
\end{abstract}


\section{Introduction}
\label{intro}
When evaluating interventions in clinical trials, time-to-event (T2E) endpoints such as overall survival (OS) or time to a severe cardiovascular event play a prominent role. Such endpoints are defined as the time between two clearly defined events, e.g. from registration into a trial or randomisation to the earliest of (a subset of) death, appearance of tumor, development of some disease, recurrence of a disease, and so forth. \te endpoints are commonly used in many indications, e.g. oncology \cite{marcus_17}, Alzheimer's and Parkinson's \cite{mcghee_16}, multiple sclerosis \cite{montalban_17}, or cardiovascular diseases \cite{mcmurray_14}.

The appearance of new types of treatments and the multiplication of lines of treatment, especially in oncology, have resulted in the use of surrogate endpoints for OS such as progression-free (PFS), or disease-free survival (DFS). Their development is strongly influenced by the necessity of reducing clinical trial duration, cost and number of patients \cite{bellera_13}. However, while these endpoints are frequently used, there is consensus in the literature (\cite{peto_77}, \cite{altman_95_surv}, \cite{mathoulin_08} \cite{bellera_13}), persistent over time, that these endpoints are often poorly defined and definitions can differ between trials. For example, Mathoulin-Pelissier et al. \cite{mathoulin_08} found that about half of articles they reviewed, and which were published in major clinical journals, failed to provide a clear definition of the \te endpoint, and 68\% reported insufficient information on the survival analysis.

In 2014, the Steering Committee of the International Conference on Harmonization (ICH) endorsed the formation of an expert working group to develop an addendum to the ICH E9 guideline ({\it Statistical Principles for Clinical Trials}) \cite{mehrotra_16}. This ICH E9 draft addendum (\cite{iche9r1}, henceforth simply referred to as ``addendum'') has been released for public consultation in August 2017. It focuses on two topics particularly relevant for confirmatory randomized clinical trials (RCT): {\it estimands} and {\it sensitivity analyses}. According to the addendum, an estimand describes what is to be estimated based on the question of interest and can be defined through the population of interest, endpoint of interest, specification of how intercurrent events are handled, and summary measure. In what follows, we refer to the specification of how intercurrent events are handled as {\it intervention effect}. A sensitivity analysis ``...can help to investigate and understand the robustness of estimates; the sensitivity of the overall conclusions to various limitations of the data, assumptions, and approaches to data analysis.'' \cite{ICHE9_conceptpaper}.

The need to amend E9 with a discussion on estimands grew out of the realization of an apparent lack of alignment between the objectives stated in a clinical trial protocol and the accompanying quantification and interpretation of the ``treatment effect'' reported in a regulatory submission. While the estimand framework has been developed with different clinical trial settings and endpoints in mind, the examples discussed in publications, at scientific meetings, and in the addendum have largely focused on symptomatic studies and continuous, longitudinal endpoints (\cite{akacha_15}, \cite{holzhauer_15}, \cite{mehrotra_16}). Given the realization of lack of common definition for \te endpoints, and the introduction of the estimand concept in the addendum, we would like to connect these two aspects and discuss estimand considerations as they apply to outcome trials that focus on \te endpoints. We hope that this, together with further efforts discussed in Section~\ref{heterogeneity}, will lead to alignment in endpoint, and ultimately, estimand definitions.

While this paper was under review, Unkel et al. \cite{unkel_18} have also looked into estimands for \te endpoints, as they apply to safety analyses with different follow-up times in treatment arms. This illustrates that considerations for T2E endpoints within the estimand framework do not only apply to efficacy, but also safety or quality-of-life \te endpoints.

The key messages we want to convey in this paper are:
\bi
\item Embed \te endpoints in the estimand framework.
\item Illustrate that {\it censoring} should be considered part of the estimator (not estimand) definition.
\item We discuss the (often implicit) assumptions made when censoring at an intercurrent event. The decision on how an estimator deals with an intercurrent event, e.g. whether it censors it or not, should be derived from the estimand which in turn has to be defined first.
\item If we have non-proportional hazards (NPH) and censoring, the estimand that is estimated by the hazard ratio from Cox regression depends on the censoring distribution. This is a rather undesirable feature and we raise the question whether, within the estimand framework, the Cox regression hazard ratio should remain the method of choice to quantify a treatment effect for \te endpoints. We discuss some potential alternatives.
\item Related to this, we raise the question whether {\it hypothesis testing} and {\it effect estimation} need to be tied to the same estimand, or whether these can be considered separately.
\item Based on several examples from the literature, we illustrate that seemingly clearly defined endpoints in clinical trials are subject to substantial heterogeneity in how they are specified. We anticipate that the estimand framework will help aligning these definitions.
\item We retrospectively embed the analysis specification of a large Phase 3 oncology RCT into the estimand framework. We hope this can serve as a basis for trial teams to specify estimands moving forward and to inform data collection strategies.
\item The addendum uses causal inference language and concepts, e.g. the principal strata strategy. However, it is not clear whether post-addendum estimands need to be causally interpretable. We would welcome clarification of this aspect in the final version of the addendum.
\ei
The paper is structured as follows: In Section~\ref{attributes} we review the estimand attributes proposed in the addendum and how they apply to \te endpoints while Section~\ref{censoring} briefly summarizes how we see the role of {\it censoring} within the estimand definition. The connection between logrank test, Cox proportional hazards regression (CPHR), and the proportional hazards (PH) assumption, especially if the latter is not met, is discussed in Section~\ref{coxproblem}. Section~\ref{heterogeneity} is devoted to a brief review of examples from the literature that illustrate how heterogeneous seemingly ``clearly defined'' endpoints actually are between trials and the implications of this heterogeneity. In Section~\ref{diffestdiffdis} we illustrate how intercurrent events can play a different role and thus necessitate a different estimand strategy depending on the indication the trial is run in. As an illustration, the endpoints and sensitivity analyses (``sensitivity'' with the pre-ICH E9 addendum meaning) of a large Phase 3 oncology trial, that has recently been reported, are then discussed in view of the new estimand framework in Section~\ref{gallium}. We conclude with a discussion in Section~\ref{discussion}.

\section{Estimand attributes for \te endpoints}
\label{attributes}

In order to provide a framework for the following sections we discuss how \te endpoints can be viewed in the estimand framework. According to the addendum, an estimand describes what is to be estimated based on the question of interest and can be defined through the population, variable, intervention effect, and summary measure. Among these, the intervention effect, which specifies how {\it intercurrent events} are reflected in the scientific question, may be considered the attribute that in general adds most novelty to the way how we describe trial objectives. Intercurrent events are clinical events that occur between randomisation and before the endpoint, such as non-adherence, discontinuation of intervention, study withdrawal, treatment switching, use of rescue medication, second line treatment, transplantation, or death. The summary measure specifies the quantity on which the treatment comparison will be based.

Note that the four attributes of an estimand as described in the addendum should not be considered independently, but in relation to each other (\cite{iche9r1}, Section A.3.1). In what follows, we discuss them as they apply to a \te endpoint.

\subsection{Population}
\label{population}
The {\it population} to be sampled from when generating clinical trial data is usually characterized through a comprehensive list of in- and exclusion criteria. We do not identify anything specific to \te endpoints concerning the definition of the population under study. However, as mentioned by a reviewer, the rate of intercurrent events in the population, as defined in the {\it intervention effect} attribute of the addendum, will have a direct bearing on the value of the estimand. As a consequence, the estimand, and thus also the estimate of interest, is becoming more dependent on the characteristics of the population. These considerations apply not only to \te but any type of endpoint.

\subsection{Variable}
\label{variable}
When defining an estimand, the attribute {\it variable} is intended to summarize the quantities required to address the scientific question. For a \te endpoint this would be, in line with Altman and Bland \cite{altman_95}:
\bi
\item Starting date: Typically, in a clinical trial this is either the date of registration into a (single-arm or non-randomized) trial or the date of randomisation. For simplicity, we will generally use ``randomisation'' to refer to the starting date.
\item Event defining the endpoint: As discussed e.g. in Chapter 2 of Klein and Moeschberger \cite{klein_03}, the event of interest may be death, appearance of tumor, development of some disease, recurrence of a disease, and so forth.
\ei
In the simplest case, e.g. when looking at OS, the starting date and event date are precisely defined up to the exact day. On the other hand, when the event date is determined through pre-specified assessments, e.g. regular tumor imaging in oncology, then the resulting \te might have a density that has peaks around these assessment dates if one simply assumes that the data is right-censored at the actual assessment date (instead of interval-censored with interval between two consecutive assessment dates). This manifests through ``steps'' in the survival curve when nonparametrically estimated through, e.g., Kaplan-Meier. An example for this pattern is PFS in the CLEOPATRA trial \cite{baselga_12}.

The definition of the endpoint might involve more than one timepoint. We discuss an example from Multiple Sclerosis in Section~\ref{heterogeneity}. Interestingly, this also implies that intercurrent events can not only happen between randomisation and the endpoint, but also between the initial and final assessment of the endpoint itself.

\subsection{Intervention effect}
\label{intervention}
The {\it intervention effect} defines how intercurrent events, i.e. potentially treatment-related clinical events that occur between randomisation and before the endpoint, are reflected in the scientific question. As a matter of fact, what constitutes an intercurrent event for a given variable depends on the variable itself and the disease indication, i.e. the population attribute. A key difficulty when defining the summary measure in Section~\ref{summary} is that an observed intercurrent event typically depends on the treatment received. We provide a list of intercurrent events in a standard oncology setting in the case study in Section~\ref{gallium}.

In what follows, we discuss the strategies proposed in the addendum (Section A.3.2) to handle intercurrent events as they apply to a \te endpoint.
\bi
\item ``Treatment policy'': The value for the variable is used regardless of whether or not the intercurrent event occurs,
\item ``Composite'': make the intercurrent event part of a composite endpoint by counting it an event defining the endpoint,
\item ``Hypothetical'': a scenario is envisaged in which the intercurrent event would not occur, e.g. because the patient switched treatment,
\item ``While on treatment'': we are interested in the response to treatment prior to the occurrence of the intercurrent event, e.g. start of second line treatment in the absence of observing the endpoint,
\item ``Principal stratum'': restrict the population of interest to the stratum of patients in which an intercurrent event would not have happened.
\ei
Note that a dedicated strategy for each intercurrent event needs to be defined, implying that in case of multiple intercurrent events, more than one of the above strategies may be needed to define a single estimand.

Now, assume our trial objective would be ``measuring the time between randomisation and progressive disease (PD)''. The variable attribute of the corresponding estimand would then be ``time from randomisation until PD'' and death would be an intercurrent event. For a \te endpoint, the intercurrent event of death can be embedded in the addendum strategies as follows:
\bi
\item If we counted death as an event we made the intercurrent event part of a composite, and we get PFS as the variable attribute of our estimand.
\item Death could be considered an intercurrent event {\it competing} with progression. In order to embed this case into the addendum language, recall the addendum definition of the ``while on treatment strategy'': ``...we are interested in the response to treatment prior to the occurrence of the intercurrent event. If a variable is measured repeatedly, its values up to the time of the intercurrent event may be considered to account for the intercurrent event, rather than the value at the same fixed timepoint for all subjects.'' So, for a longitudinal endpoint, which is referred to in this definition, we would simply impute the value of the variable (e.g. numerical score that defines the variable) at the intercurrent event (when patient dies) as the value at the fixed timepoint. We propose to fit a competing intercurrent event for a \te endpoint in the ``while on treatment strategy'' of the addendum, as we are equally interested in the response to treatment prior to either the event of interest or the competing event. On the estimator level this would mean we would simply censor at death when performing inference on time-to-progression (TTP). As discussed in Section~\ref{censoring} one then needs to make sure to align the summary measure with this censoring strategy.
\item If we are willing to entertain the assumption that those who die have the same momentary risk of an event as those that remain in the risk set, we then estimate with this the hypothetical estimand ``time from randomisation until PD, assuming that the time until PD of patients who died is imputed using the longer term outcomes of other patients who survived and remained under observation'' \cite{fleming_09}, and we retrieve the familiar definition of TTP \cite{fda_endpoints}, \cite{pazdur_08}. The assumption e.g. holds in case of {\it random censoring}, i.e. if the censoring time and \te are stochastically independent \cite{allignol_16}.
\ei
So, interestingly, both the hypothetical and the ``while on treatment'' strategy can be estimated through censoring at the intercurrent event of death. However, the estimand the resulting estimators are targeting is different, and this manifests itself in the summary measure attribute of the estimand. We revisit a few points around censoring and summary measures in the case of competing risks in Section~\ref{censoring}.

As becomes clear, the choice between TTP, a competing risk analysis, and PFS is ultimately a decision about which estimand to look at, and this estimand should be inferred from the trial objective. Note that compared with TTP, PFS is generally the preferred regulatory endpoint \cite{fda_endpoints}.

The PFS example can also be used to discuss the general concept of a treatment policy estimand, where the following considerations apply to any type of endpoint, not just time-to-event. Treatment policy is defined by encompassing the intercurrent event. The rationale for doing so is that if the intercurrent event impacts the endpoint, this will be reflected in the resulting treatment effect, i.e. the causal link is preserved. This boils down to ``ignoring'', or not censoring at, the intercurrent event, and that is often considered the ``analogue'' to the ``intention-to-treat'' (ITT) principle introduced in the original ICH E9 guideline, although much ambiguity is involved around what is considered to constitute the ITT principle \cite{leuchs_17}. We would like to caution against the use of such a broad strategy when defining a treatment policy estimand. If any intercurrent event, whether pre-specified in the protocol or not, is ignored and a treatment policy estimand is postulated, the implied estimand is difficult to interpret. This is because allowing any intercurrent event is not really a strategy and causality remains unclear. We thus recommend that any intercurrent event that is ignored for a treatment policy strategy should still be pre-specified in the protocol and be systematically collected, somehow leading to a ``protocol-defined treatment policy''. Related, and also not specific to \te endpoints, Hernan and Robins address issues that may arise from uncritical reliance on the intention-to-treat principle in pragmatic trials \cite{hernan_17nejm}. They provide guidance on how per-protocol analyses can be used to overcome these issues.

Finally, if applicable, a principal stratum strategy resolves the problem that the occurrence of an intercurrent event is generally associated to treatment. This advantage of the principal stratum strategy comes at the cost of the difficulty to identify the patients in each stratum. This membership can typically only be inferred through suitable covariates, unless the intercurrent event is completely unrelated to treatment, but then the ``random censoring'' assumption would also be met, making a hypothetical estimand straightforward to estimate. An example is discussed by Shepherd et al \cite{shepherd_07} and an estimator of that estimand is proposed by Chiba \cite{chiba_13}. An introduction into principal stratification and how it allows for causal inference is provided by Rubin \cite{rubin_06}. Rubin discusses the issues that come with intercurrent events using the term ``intermediate outcome variables'' for intercurrent events, and the concepts are not just limited to \te endpoints. He specifically looks into ``truncation by death'', so precisely the scenario we discuss above.

In general, the choice made for {\it each} potential intercurrent event for a \te endpoint, i.e. making it part of a composite, treat it as competing, be interested in a hypothetical world where it would not happen, ignore it, or apply the principal strata strategy, requires intensive discussions about the actual precise trial objective and the implied estimand.

We recommend to document these choices and the resulting estimands, as well as the data analysis strategy, in trial protocols, statistical analysis plans (SAP), and publications, e.g. in tabular form similar to Table 2 in Bellera et al. \cite{bellera_15}.

For illustration, we provide an oncology case study in Section~\ref{gallium}.

\subsection{Summary measure}
\label{summary}
In this section, we first discuss our view on the connection between {\it statistical hypothesis testing} and {\it effect quantification} in the estimand context. The addendum specifies in Section~A.1. as part of its scope that it ``...presents a structured framework to link trial objectives to a suitable trial design and tools for estimation and hypothesis testing.'' However, in what follows the addendum does not explicitly discuss the connection between {\it testing} and {\it estimation}, and the paragraph on the summary measure clearly focuses on estimation, not testing. This may be because for the endpoint type that initiated the addendum, continuous data collected over time with intercurrent events, the statistic used for testing and the effect estimate are closely linked and allow for a consistent interpretation. Depending on how the population survival functions relate to each other, this link may break down for the most commonly used methods to analyze \te endpoints, namely the logrank test and CPHR: If we have non-proportional hazards, the estimand implied by the estimate based on CPHR depends on the censoring distribution, and might thus vary from one trial to the next \cite{boyd_12, aalen_15}. Furthermore, since the logrank test is intimately connected with CPHR, the estimand implied by the logrank test suffers from the same deficiency. We will provide a discussion of these aspects in Section~\ref{coxproblem}. The potential disconnect between testing and estimation in certain cases offers two distinct ways of making the summary measure component of the estimand more specific:
\begin{enumerate}
\item Either one insists that testing and estimation are consistent. The implication of this approach would be that for \te endpoints, methods such as logrank test and CPHR would potentially need to be replaced by alternative tests and estimators that are robust against violation of the PH assumption, see Section~\ref{coxproblem}.
\item Or one allows for a two-stage procedure: In a first stage, a valid hypothesis test is performed which serves as a gatekeeper. ``Valid'' always refers to maintaining type I error. If the null hypothesis under consideration is rejected and the effect estimate points in the right direction, then the trial is considered a success. The analysis then goes on in a second stage to quantify the effect by which the experimental treatment is ``better'' than the control. This effect quantifier would not necessarily have to be connected to the test statistic of the gatekeeper test.
\end{enumerate}
Based on our experience with several Health Authorities (HA), once the null hypothesis has been rejected using a valid hypothesis test, it is possible to choose one or more summary measures which might differ from the one that corresponds to the performed hypothesis test, even if the chosen summary measure would not reject its associated null hypothesis. Akacha et al. \cite{akacha_17} briefly discuss this aspect in Section 5.7 as well. So it appears that at least implicitly, the second approach above is acceptable to HAs.

In what follows, we discuss effect quantifiers for a \te endpoint that are simple, well-known, correspond to a clear estimand, and can easily be computed from available data. A discussion of alternatives to the hazard ratio under more general assumptions is provided in Section~\ref{coxproblem}.

Recall that in a two-arm RCT with a \te endpoint, the necessary number of events is typically determined such that a two-sided logrank test of the null hypothesis
\begin{center}
$H_0: \ \ S_\text{control} \ = \ S_\text{experimental}$
\end{center}
has the pre-specified power for an assumed alternative hypothesis, if all assumptions are met \cite{cook_08}. Here, $S_i = P(T_i > t)$ is the survival function in treatment arm $i \in $ \{control, treatment\}, with $T_i$ the corresponding survival time, a non-negative random variable, and $t \ge 0$.
If $H_0$ is rejected using a valid test and the effect estimate points in the right direction, the trial is considered a success and the effect has to be quantified.

One candidate to use is an estimate of the difference $(S_\text{experimental} - S_\text{control})(t_0)$ at some timepoint $t_0$ (``milestone survival'') as our summary measure after passing the gatekeeper. Further potential effect quantifiers that describe and summarize the observed data are median or any other quantile difference, or difference in restricted mean survival time \cite{royston_13}, the latter being especially relevant in pharmacoeconomic applications, see e.g. \cite{demiris_15}. Instead of differences, ratios or even odds ratios can be based on $S_\text{experimental}$ and $S_\text{control}$.
All these estimands rely on an estimate of the survival function, such as Kaplan-Meier, or one based on a parametric assumption. The assumptions for the chosen estimator, e.g. about censoring and the suitability of the parametric model, need to be checked and any summary measure should be accompanied by a quantification of uncertainty, most likely a confidence interval. Whether and how some of these estimands are amenable to a causal interpretation is discussed in \cite{mao_18}.

The example discussed in Section~\ref{diffestdiffdis} allows for an illustration of the interplay between {\it intercurrent event} and {\it summary measure} in general. We thus defer the discussion on this aspect of the summary measure to that section.

\section{Censoring}
\label{censoring}

When developing a clinical trial, in line with the recommendations by the National Research Council \cite{nrc_10}, the order of what needs to be discussed is
\begin{enumerate}
\item trial objective,
\item estimand,
\item study design,
\item data collection and handling strategy,
\item estimator.
\end{enumerate}
As a reviewer pointed out, if an intercurrent event is not terminal, then one may choose to {\it truncate} the T2E variable, i.e. the estimand, at this intercurrent event, and this might often be referred to as ``censoring''. However, we propose to leave the word ``censoring'' to activities related to the estimator, and use ``truncation'' instead for the estimand definition.

Censoring is thus part of the estimator definition and describes how to handle data that is only partially observed for the \te variable, either because an intercurrent event happened before the event defining the variable or the trial ended.

The event {\it end of trial} plays a special role in the context of censoring. Note that with ``end of trial'' we imply analysis at any pre-specified clinical cutoff, e.g. also when performing an interim analysis. This cutoff is typically pre-specified at baseline \cite{hernan_17}, which implies that the time until the clinical cutoff date is definitely independent of the time-to-event, i.e. we have random censoring. Using an estimator based on simple right-censoring, generally referred to as {\it administrative censoring}, is thus uncritical \cite{allignol_16}.

An estimator involving censoring at an intercurrent can be foreseen for two of the five strategies described in Section~\ref{intervention}:
\bi
\item If the estimand specifies a ``while on treatment strategy'' we consider the situation of an event of interest and a competing event. It is important that in this case, the estimator is aligned with the summary measure attribute of the estimand: hazard-based inference is unbiased for each event-specific hazard separately when censoring at the intercurrent event(s), see e.g. Beyersmann et al. \cite{beyersmann_12}. A simple Kaplan-Meier type estimator for the survival curve of the event of interest that simply censors at the competing intercurrent event(s) is biased though \cite{allignol_16}. To get probability statements, one has to look at cumulative incidence functions, or risk quantifiers based on subdistribution hazards \cite{fine_99}. We refer to Geskus \cite{geskus_15} and Unkel et al. \cite{unkel_18} for a discussion of these aspects.
\item If the estimand specifies a hypothetical strategy, simple right-censoring at the intercurrent event provides unbiased estimates if we e.g. assume {\it random censoring}, as discussed in Section~\ref{intervention}.
\ei
Andersen \cite{andersen_05} and Allignol et al. \cite{allignol_16} discuss the subtle difference between {\it random} and {\it independent} censoring. Andersen explains that precise mathematical formulations of independent censoring may be given and that frequently used models for the generation of right-censored satisfy these conditions. However, the conditions may be impossible to verify for actual data. Interpreted within the addendum discussion, this re-iterates the need for sensitivity analyses when assumptions on censoring mechanisms, and thus estimators to estimate the above estimands, are specified, as also pointed out by Unkel et al. \cite{unkel_18}.

\section{Logrank test, Cox proportional hazards regression, and estimating the estimand}
\label{coxproblem}
For \te endpoints, the logrank test is overwhelmingly used as a gatekeeper to test $H_0$. For the logrank test to be valid, the following assumptions about the data are made: independent censoring, the survival probabilities are the same for subjects recruited early and late in the trial, and the events happened at the times specified \cite{bland_04}. While its power is maximal if the hazard functions corresponding to the assumed survival curves are indeed proportional, the logrank test maintains type I error, and is thus valid, also under NPH. This, because under $H_0$ we assume identical survival curves and the PH assumption does trivially hold. The PH assumption is only used to define the alternative and thus the nature of the test statistic.
All this justifies the important role of the logrank test as a gatekeeper for testing $H_0$ for a \te endpoint in a regulatory context.

However, while the logrank test can be developed as a hypothesis test comparing the standardized difference between expected and observed number of events, the fact that it corresponds to the score test in a CPHR, see e.g. Section 3.9 in Collett \cite{collett_03}, implies that the estimand connected to the logrank test is the same as the one for CPHR, i.e. relies on the PH assumption if we deviate from $H_0$. So, under NPH, it is a priori not clear to what estimand the logrank test corresponds.

Now, under NPH and if there is no censoring, Xu and O'Quigley \cite{xu_00} derive that the estimate from a CPHR can still be interpreted as a regression effect suitable averaged over time, i.e. has an interpretable estimand. However, if there is right-censoring, already Struthers and Kalbfleisch \cite{struthers_86} have shown that the estimand targeted by CPHR is defined implicitly through an estimating equation that depends on the censoring distribution. As discussed in Nguyen and Gillen \cite{nguyen_12}, this implies that in the presence of right censoring, the estimand that is estimated by CPHR under NPH depends on the censoring pattern of the actual data, which leads to an estimand that varies with the censoring distribution. The dependence of the {\it estimand value} on the censoring distribution is nicely illustrated by Nguyen and Gillen (\cite{nguyen_12}, Figure 2). As a consequence, the estimand targeted by CPHR becomes trial-specific, as it is virtually impossible to replicate the censoring distribution from one trial to the next, even if only administrative censoring is present. This, because the censoring distribution at least depends on the accrual pattern and the trial length, which are typically outside of the trial sponsor's control. Or, as Boyd et al. \cite{boyd_12} put it, ``...the usual unweighted Cox estimator will be consistent for a parameter that is dependent upon patient accrual and dropout patterns that bear no relevance to the scientific objectives of a clinical trial.'' This quote nicely relates to the estimand discussion, as the primary goal of the addendum is to align scientific objective, estimand, and estimator.

As discussed in Section~\ref{intervention}, even in the absence of intercurrent events, we typically have to deal with at least administrative censoring in a trial with a \te endpoint. Xu and O'Quigley \cite{xu_00} and Boyd and et al. \cite{boyd_12} provide estimators of an average regression effect for the continuous time CPHR model, the latter under more general assumptions on censoring. Extending the work of Struthers and Kalbfleisch \cite{struthers_86}, these estimators target an estimand that is again only implicitly defined. However, Xu and O'Quigley \cite{xu_00} nicely show that the estimator approximates the population average effect $\int \beta(t) d \, F(t)$, where $\beta(t)$ is the time-varying regression coefficient, and $F$ is the marginal distribution of the failure times. Nguyen and Gillen \cite{nguyen_12} provide similar results for a discrete hazard model. The key properties of these estimators are that they are equal to the one from CPHR under PH (\cite{xu_00}, Section 5), but they are robust against the censoring distribution. This means they are consistent for the same estimand as CPHR if the PH assumptions is true, and they estimate a quantity that is independent of the censoring distribution in case of NPH.

In conclusion, by replacing the logrank test and the estimator based on the CPHR by e.g. the methods introduced by Xu and O'Quigley \cite{xu_00} or Boyd et al. \cite{boyd_12}, it would be possible to remove the conceptual problems discussed above, i.e. one would get a hypothesis test and effect quantifier that
\bi
\item have a clear estimand
\item which would be (asymptotically) independent of the censoring distribution
\ei independently of the PH assumption.

\section{Heterogeneity in \te endpoint definitions}
\label{heterogeneity}

In this section we focus on heterogeneity of endpoint definitions across studies, and illustrate the implications of this heterogeneity.

In the terminology of the addendum, estimand heterogeneity for a \te endpoint primarily affects the interplay between {\it intervention effect} and {\it variable}, i.e. which clinical events are considered for one or the other, or ignored, make the endpoint incompletely observed, or treated as competing. The lack of common definitions for \te endpoints has previously been recognized in the literature, see Peto et al. \cite{peto_77} for the earliest account to the best of our knowledge, including recommendations how to define and report \te endpoints. Still, the review by Altman and Bland \cite{altman_95_surv} finds frequent failure to specify whether non-cancer deaths were treated as events or censored, or how deaths without relapse were considered in the definition of the endpoint. They re-iterate the recommendation to give a clear definition of the ``time origin, the event of interest and the circumstances where survival times are censored'' for each endpoint considered. More than another decade later, Mathoulin-Pelissier et al. still found that about half of the reviewed articles published in major clinical journals failed to provide a clear definition of the \te endpoint, and 68\% reported insufficient information on the survival analysis \cite{mathoulin_08}.

Bellera et al. state that ``Most of these \te endpoints currently lack standardised definition enabling a cross comparison of results from different clinical trials'' \cite{bellera_13}, clearly pointing to the need of a discussion of estimands also in this context, although the authors of the latter paper do not use the term. Even for a binary endpoint, Kahan and Jairath \cite{kahan_18} also find marked differences in estimated odds ratios, depending on endpoint definition.

An example illustrating the lack of a clear endpoint definition is discussed by Birgisson et al. \cite{birgisson_11}. For colorectal cancer, these authors find that the inclusion of second primary cancer as an event in the definition of DFS relevantly alters effect estimates. They recommend that researchers and journals must clearly define \te endpoints in all trial protocols and published manuscripts. A similar analysis has been performed for colon cancer \cite{punt_07}.

DFS is also the primary end point for many large adjuvant breast cancer trials. The results of such trials, if positive, will likely change clinical practice, refer to Minckwitz et al. \cite{minckwitz_17} for a recent example. Historically, which of the five addendum strategies is applied to a given clinical event has been inconsistent. The typical definition involved local, regional, and distant recurrence of the tumor, as well as death as events for the variable, DFS. Often inconsistently handled were
\bi
\item initiation of treatment for contralateral breast cancer,
\item second primary cancers, further depending on whether they were contralateral or nonbreast,
\item and death not due to breast cancer.
\ei Hudis et al. (\cite{hudis_07}, Table 1) gives an overview and the paper provides recommendations reached by an expert group. Building on this, the DATECAN initiative (see below, \cite{gourgou_15}) applied a formal international consensus method, to increase the use and acceptability in current practice of common definitions. This illustrates that work is still necessary and is being done to align the definition of the variable attribute in the estimand context. We hope that with the addendum this alignment will also carry over to the other estimand attributes.

The aim of the {\it Definition for the Assessment of Time-to-event Endpoints in CANcer trials} (DATECAN) initiative is to provide recommendations for standardised definitions of \te endpoints \cite{bellera_13} not only for breast but for many cancer indications: those for pancreatic \cite{bonnetain_14}, sarcomas and gastrointestinal stromal \cite{bellera_15}, breast \cite{gourgou_15}, and renal \cite{kranar_15} tumors have been published so far. Note that all these efforts are currently made without explicit referencing to the estimand framework in general, or the framework put forward by the addendum. The consensus defines how intercurrent events and the variable are recommended to be handled, but the other estimand attributes, {\it population} and {\it summary measure}, are left open.

The Steering Committee of the DATECAN project considered defining censoring rules to be a statistical rather than a clinical question \cite{kranar_15}. As a result, censoring rules were not discussed during the DATECAN consensus process. Disentangling the estimand from the estimator is in line with our recommendation in Section~\ref{censoring}, i.e. the clinical question is the definition of the estimand and it remains then a statistical challenge to estimate the targeted quantity. However, from the description in Bellera et al. \cite{bellera_13}, it remains unclear how much the decisions on the actual estimand made in the DATECAN initiative are actually derived from a clearly defined scientific objective.

Another initiative with the goal of improving reporting of clinical trials is the CONsolidated Standards Of Reporting Trials (CONSORT) statement \cite{consort_bmj}. Although CONSORT is aimed at structuring reporting of RCTs in general, many of the items on the CONSORT checklist aim at improving the definition and description of trial objectives, population, endpoints, and {\it outcomes and estimation}. So, the CONSORT statement actually covers a substantial portion of the four attributes that make up an estimand in the addendum, for any type of endpoint, not only time-to-event. The addendum can thus be seen as detailing the aspects pertaining to these CONSORT items further, and connecting them more directly to the trial objective.

Acknowledging the current state of heterogeneity in the definition of \te endpoints, in what follows we would like to point out what this potentially implies in RCTs.

As discussed in Section~\ref{summary}, whether an RCT is considered a ``success'' generally depends on rejection or non-rejection of the considered null hypothesis. So the question of ``statistical significance'' carries substantial weight, and pronouncedly so for trials potentially leading to regulatory approval. Montalban et al. \cite{montalban_17} provide an example for Multiple Sclerosis: The primary variable was time-to-initial disability progression (time-to-IDP), which had to be confirmed 12 weeks after IDP, and the associated logrank test was statistically significant. However, since not all patients come back 12 weeks after their initial assessment to have their progression confirmed, for the primary analysis it was assumed that all these patients progressed (Table S10 in \cite{montalban_17}). An alternative definition of the endpoint, imputing half of the patients randomly as event and the other half as censored, was analyzed as part of a range of sensitivity analyses, yielding a non-significant logrank test. This example also provides a case study of a variable attribute that is assessed at different timepoints, implying that intercurrent events can also happen within the variable attribute itself.

Similarly, van Cutsem et al. discuss colon cancer, with endpoint DFS in the PETACC-3 trial \cite{vancutsem_05}. The primary definition in the trial protocol counted second primary cancer other than colon as an event, and with this definition, the trial is reported to be ``non-significant'' \cite{vancutsem_05}. However, the evaluation of relapse-free survival (RFS) as defined in the PETACC-3 protocol (which corresponds to the definition of {\it DFS} in the MOSAIC trial, \cite{andre_04}!) showed a $p$-value of 0.02 and was called ``statistically significant'' \cite{vancutsem_05}. 

Not surprisingly, estimates of event-free probabilities at 3 and 5 years are different for different variable definitions as well \cite{vancutsem_09}, see Bellera et al. for further details \cite{bellera_13}. Note that the trial objective of ``improving DFS'' is not nearly specific enough to imply unambiguous definitions.

Meta-analyses will not provide reliable summary estimates if studies with different definitions of variables and/or intercurrent events are being compared \cite{birgisson_11}.

``Dynamic borrowing'' of historical data has gained some popularity recently \cite{viele_14}, but that also raises the question of ``compatibility'' of the historical data used in these kind of trials, not only from a statistical perspective but also in terms of estimand definition.

In an analysis combining data from several trials in colon cancer to assess surrogacy of DFS for OS after various times of follow-up \cite{sargent_11}, among them PETACC-3 \cite{vancutsem_09} and MOSAIC \cite{andre_04} for which DFS definition was different in terms of handling second tumors \cite{vancutsem_05}, the DFS definition given is: ``DFS was defined as the time from randomisation to the first event of either disease recurrence or death due to any cause.'' No discussion of how second tumors were handled can be found in this surrogacy assessment, thus the reader is left uncertain which estimand was actually targeted by the analysis. The authors further found that the previously established surrogacy of DFS after 2- and 3-year median follow-up for OS in Sargent et al. \cite{sargent_05} was now modest to poor in this follow-up analysis based on six new trials not included in the previous analysis. The authors attribute this reduction in the amount of surrogacy based on the same amount of follow-up (2 and 3 years for DFS, 5 and 6 years for OS) as in Sargent et al. \cite{sargent_05} to generally longer DFS and OS achieved in colon cancer over time through improved therapy and combinations. One can only speculate on the possible contribution of heterogeneity in DFS variable definition to the lack of surrogacy in this case.

Effect estimates depending on the definition of an estimand may also complicate planning of future trials, as assumptions for the control arm in a new trial are often based on estimates of the treatment arm in previous trials.

To summarize, unprecise definitions may impact statistical significance, effect estimates, feasibility of meta-analysis, dynamic borrowing, and surrogacy assessments, and can bias assumptions for planning of future trials.

The addendum (Section A.4) emphasizes the need for construction of a suitable estimand when summarising data across trials. This means that to allow cross-trial summaries, the estimand in each single trial needs to be described in sufficient detail, to enable an assessment of which studies to combine and to quantify the ``estimand heterogeneity''.

\section{Analysis strategy for given intercurrent event depending on indication}
\label{diffestdiffdis}

In this section, we illustrate how the same intercurrent event, initiation of new therapy in lymphoma in the absence of PD, might need to be considered differently for the same \te endpoint, depending on the disease indication considered.

PFS is a commonly accepted regulatory endpoint in Non-Hodgkin Lymphoma (NHL), for both, the indolent (follicular, FL) and aggressive (diffuse large B-cell, DLBCL) subtype \cite{cheson_07}. FL is incurable and DLBCL is potentially curable, but patients failing to achieve a complete remission (CR) after initial induction treatment have a dismal outcome.
This means that physicians might treat such a DLBCL patient with a new anti-lymphoma therapy (NALT) of their choice already {\it prior} to formal imaging-based determination of progression to the first-line therapy in case of non-CR (i.e. if they have either stable disease or partial response), in order to maximize the odds to still get the patient to CR. So, NALT can be considered a potential intercurrent event for the endpoint of PFS. For FL, in line with regulatory guidelines \cite{fda_endpoints}, PFS is generally defined as time from randomisation to the earlier of disease progression or death \cite{marcus_17}, and NALT is ignored for PFS, implying a treatment-policy estimand. Ideally, those NALTs that are allowed to be administered within the trial are protocol-specified, see the discussion in Section~\ref{intervention}.

In DLBCL, PFS is often defined as for FL (e.g. in the GOYA trial, \cite{vitolo_17}), although the lymphoma-specific harmonization effort by Cheson et al. \cite{cheson_07} indicates that ``...in studies in which failure to respond without progression is considered an indication for another therapy, such patients should be censored at that point for the progression analysis.'' The estimand corresponding to the definition of the variable PFS in this case would be a hypothetical one. The censoring rule implies that for patients who fail to respond, their PFS is imputed (terminology used by Fleming et al., \cite{fleming_09}) based on patients that did respond, but had longer follow-up. Interestingly, in another highly cited paper it is recommended that ``...patients should not be censored at the time other treatments are initiated when analyzing the PFS end point...'' \cite{fleming_09}. Here, the interest focuses on a treatment-policy estimand strategy. Given the conflicting recommendations given by regulatory guidelines and key opinion leaders, how should a researcher designing a trial using PFS as the primary endpoint proceed to define this endpoint? In the framework of estimands for a \te endpoint outlined in Section~\ref{attributes}, the definition of the variable that defines PFS appears unambiguously accepted: starting date is randomisation and the event defining the endpoint is the earlier of PD and death. To define the potential estimand we identify two potential intercurrent events:
\bi
\item Failure to respond to induction therapy, but no PD (IC1).
\item Initiation of NALT, but no PD (IC2).
\ei

Note that it is not entirely clear from the statement in Cheson et al. \cite{cheson_07} whether these authors really think of two different intercurrent events when they write ``...failure to respond'' or whether they implicitly assume IC1 = IC2. In any case, to be as precise as possible, we consider both these events separately in our discussion, as it is not entirely implausible that a patient with partial response only at end of induction will not immediately receive NALT. A number of estimands that could be defined by treating IC1 and IC2 are summarized in Table~\ref{PFSestimands}. For the estimation of the estimands in Table~\ref{PFSestimands} we propose to administratively censor at the last non-PD tumor assessment for the clinical cutoff, and also to censor at the intercurrent event whenever we consider a hypothetical estimand.

As for the summary measure, the choice depends on which of the two approaches described in Section~\ref{summary} is preferred and whether the PH assumption is sensible in this indication. The preferred approach will likely evolve after the addendum has been finalized and substantive knowledge of the DLBCL indication is needed to decide on the plausibility of the PH assumption. Whether the PH assumption seems appropriate or not depends on the choices made for the other three estimand attributes, most specifically the intervention effect. For simplicity, in Table~\ref{PFSestimands} we make the PH assumption.

\begin{landscape}
\begin{table}[h]
\begin{center}
{\footnotesize
\begin{tabular}{p{2cm}|p{2.5cm}|p{1.8cm}|p{7cm}|p{3.5cm}|p{5.5cm}}
  \hline
  Estimand & Population & Variable & Intervention effect & Summary measure & Comment \\ \hline
  Option 1 & DLBCL patients, defined by list of in- and exclusion criteria & PFS & NALT: {\bf treatment policy} \newline Failure to respond: {\bf treatment policy} \newline Death: {\bf composite} & logrank test and hazard ratio & Corresponds to definition in Fleming et al. \cite{fleming_09}. Used in large RCT \cite{vitolo_17}. \\ \hline
  Option 2 & as Option 1 & as Option 1 & NALT: {\bf hypothetical} \newline Failure to respond: {\bf treatment policy} \newline Death: {\bf composite} & as Option 1 & \\ \hline
  Option 3 & as Option 1 & as Option 1 & NALT: {\bf treatment policy} \newline Failure to respond: {\bf hypothetical} \newline Death: {\bf composite} & as Option 1 & Corresponds to definition in Cheson et al. \cite{cheson_07}. \\ \hline
  Option 4 & as Option 1 & as Option 1 & NALT: {\bf hypothetical} \newline Failure to respond: {\bf hypothetical} \newline Death: {\bf composite} & as Option 1 & \\ \hline
  Option 5 \newline (Event-free survival, EFS) & as Option 1 & PFS becomes EFS & NALT: {\bf composite} \newline Failure to respond: {\bf hypothetical} \newline Death: {\bf composite} & as Option 1 & Similar to PFS, may be useful in evaluation of highly toxic therapies, only acceptable for HAs if NALT is supported by some ``objective'' evaluation of treatment failure in the absence of PD. Used as part of a surrogacy assessment \cite{maurer_14}. \\ \hline
\end{tabular}
}
\caption{Potential PFS estimands for DLBCL, depending on how intercurrent events are considered. Estimand strategies according to the addendum are emphasized in {\bf bold}.}
\label{PFSestimands}
\end{center}
\end{table}
\end{landscape}

Not all estimands in Table~\ref{PFSestimands} may be of practical interest. Furthermore, Options 2-5 all contain at least one intercurrent event for which a hypothetical estimand strategy is proposed. For an estimator based on simple censoring at IC1 and/or IC2 to be unbiased, assumptions on the censoring mechanism, e.g. random or independent censoring, have to be made, see Section~\ref{censoring}.
The need to adjust for the potential dependence of the effect estimate on the intercurrent events via methods that can deal with, what they call, ``informative censoring'' led Fleming et al. to actually advocate Option 1 in this context \cite{fleming_09}. However, the estimand framework in the addendum might potentially reverse that thinking: ideally, one first determines a trial objective which is then translated in estimand strategies to handle the intercurrent events. The last step is then to define an estimation strategy that is able to estimate the selected estimand, potentially at the price of added assumptions or the need for advanced methodology. In any case, we find that depending on how we consider the intercurrent events IC1 and IC2, we get different estimands for an endpoint that is commonly called ``PFS'', and with Option 5 even one that is considered a different endpoint. EFS has lately been considered an endpoint for DLBCL as well \cite{maurer_14}.

For FL, the situation is different, as in this indolent disease, failure to achieve response after induction therapy does not necessarily trigger NALT. Rather, physicians indeed wait with inducing NALT until the patient experiences progression, so that PFS as an endpoint is indeed more reflective of the actual clinical treatment of this disease.

The comparison between DLBCL and FL illustrates that even for two quite related diseases, for which often the same treatment is applied, defining a relevant estimand might not be straightforward and needs careful assessment on what actually a therapy under study should achieve, and what the precise trial objective is. Irrespective of which estimand is chosen, we strongly recommend that trial developers try to identify all potential intercurrent events upfront, and clearly indicate for the finally chosen estimand and each intercurrent event whether the latter is considered within a treatment policy, in the context of a hypothetical estimand, as part of a composite endpoint, or as competing. We recommend to make that transparent in a table similar to Table~\ref{PFSestimands}.

To conclude this section, and to finish the discussion on summary measure in Section~\ref{summary}, we would like to comment on the interplay between {\it intervention effect} and {\it summary measure}. The above debate on how to treat the intercurrent event ``initiation of NALT'' for PFS was settled by a general recommendation \cite{fleming_09} to use a ``treatment policy'' strategy for this intercurrent event. One, if not the major, justification to favour the treatment policy estimand is to avoid having to censor at NALT, i.e. the insight that a hypothetical estimand would potentially be difficult to estimate. With this approach, the question of NALT is dealt with in the definition of the intervention effect. Now, instead of changing the estimand from hypothetical to treatment policy to avoid having to deal with assumptions for censoring when defining the estimator, one could equally well consider adjusting the summary measure, e.g. by applying {\it inverse probability of censoring weighting} (IPCW), see e.g. \cite{robbins_93} or \cite{watkins_13}.
This method allows, again under some but now different assumptions (no unmeasured confounders, \cite{robins_97}), to estimate the survival function and consequently the treatment effect {\it if no patient had started NALT}, so again a {\it hypothetical estimand} strategy. Alternatively, an estimand defined through a rank-preserving structural failure time model (RPSFT, \cite{robins_91}) assesses the counterfactual event time of a patient, i.e. again the \te if no NALT were received \cite{watkins_13}. This can equally well be considered a hypothetical estimand. These considerations illustrate that various estimators can be defined for the same estimand. Interestingly, methodology to estimate a hypothetical estimand such as IPCW and RPSFT, have been developed long ago in the context of {\it causal inference} and successfully applied in many instances, especially in epidemiological applications. We anticipate that the addendum provides a framework and a common language that will facilitate and foster implementation of such advanced estimation methodology, because these are made necessary by the trial objective and the estimand derived from it. Furthermore, it is important to note that the ``no unmeasured confounder'' assumption for IPCW requires that (as much as possible) data on prognostic factors that explain initiation of NALT needs to be collected, by pre-specifying these factors in the {\it data collection and handling strategy}, according to the list in the National Research Council's report \cite{nrc_10}, see Section~\ref{intervention}.
Having a hypothetical estimand defined upfront will shift focus of a sponsor to collecting necessary information during the trial, another advantage of the estimand framework, as outlined in the addendum (Section A.1). Already Watkins et al. \cite{watkins_13} emphasize that ``...Often, only limited data are collected after the patient experiences the primary regulatory endpoint, which can mean important time-varying confounders are missed or that switch dates or time on/off treatment cannot be accurately defined. Careful upfront planning is required.'' We expect such planning to become more standard after the addendum is in place.

\section{Case study: the GALLIUM trial}
\label{gallium}

The GALLIUM trial \cite{marcus_17} assessed whether replacing the anti-CD20 antibody rituximab by a second generation anti-CD20 antibody, obinutuzumab, increases PFS. The trial was unblinded after it crossed the pre-specified significance level at a pre-planned efficacy interim analysis and was fully analyzed. The trial randomized 1202 FL patients and an additional 195 marginal-zone lymphoma (MZL) patients. The latter cohort can be considered a Phase 2 trial within the same protocol, whereas the first constituted the primary analysis population. Due to regional heterogeneity in standard of care, trial sites had to select one of three chemotherapy backbone therapies. Chemotherapy as well as {\it Follicular Lymphoma International Prognostic Index} (FLIPI1, \cite{solal_04}), a prognostic score used in FL, were used as stratification factors in the analysis.

Table~\ref{gallium1} provides details on the primary analysis and a list of all sensitivity analyses (with term ``sensitivity'' referring to the pre-addendum meaning in the GALLIUM protocol) reported in the clinical study report (CSR) for the primary endpoint, PFS as assessed by the investigator (Inv-PFS). We provide columns for each of the four estimand attributes introduced in the addendum. As almost all the analyses in Table~\ref{gallium1}, and also Table~\ref{gallium2}, are targeting different estimands than the primary by varying at least one of the four attributes, in the post-addendum language these are in fact all {\it supplementary} analyses. The only exceptions are variations of the primary analysis by considering an unstratified instead of a stratified logrank test and a re-randomisation test. These truly vary the underlying assumptions of the primary estimand. We have still added these variations to the summary measure column for simplicity, but they target the primary estimand, as indicated in the first column of the tables.

Depending on the endpoint, the intercurrent events we consider are death, NALT, progression, withdrawal, drop-out, discontinuation of trial treatment, and missed scheduled response assessment. While all these were systematically collected on the electronic case report form, the list of ``allowed'' NALTs was not pre-specified in the protocol, because NALT was not foreseen prior to observing the actual endpoint, PFS. So, as discussed in Section~\ref{intervention}, one can argue that it is not entirely clear what precise objective, or effect quantification, corresponds to a treatment policy estimand strategy applied to the intercurrent event NALT. However, for illustrative purposes we ignore this aspect in our case study, i.e. if in the tables below we discuss ignoring NALT for an estimand we call this treatment policy.

Table~\ref{gallium2} then continues to describe all the \te endpoints listed on \url{https://clinicaltrials.gov/ct2/show/NCT01332968}. In the original protocol, SAP, and CSR, these analyses were provided as a long list and have been selected based on substantive considerations, regulatory guidelines \cite{fda_endpoints}, experiences from previous trials, and feedback from HAs on studies in the same development program. In the estimand framework, these are {\it supplementary analyses}. We would like to emphasize that this is an {\it after-the-fact} exercise with the aim of learning how to structure an estimand and analysis plan in the future based on the addendum.

Note that the primary and some of the secondary \te endpoints were evaluated in the FL and overall, i.e. FL + MZL, population. For simplicity, we focus on the FL population only in both Tables. Adding the overall population to the lists would be a simple variation of the population attribute.

The estimand definitions in Table~\ref{gallium1} are complemented by the following data analysis strategy:
\bi
\item Clinical cutoff for PFS: administratively censor at the last non-PD tumor assessment.
\item Primary: PDs were collected after NALT, so the treatment policy strategy could be applied. However, data on PD was not routinely collected after drop-out and withdrawal. As a consequence,  estimation was based on censoring at these intercurrent events, implying a hypothetical estimand.
\item Supplementary 1: Withdrawals prior to PD: consider event at next scheduled disease assessment date in obinutuzumab arm, censored at last disease assessment for rituximab.
\item Supplementary 2: Missed assessment prior to PD or clinical cutoff date: considered event at day after last response assessment.
\item Supplementary 3: NALT prior to PD: censor at NALT.
\item Supplementary 4: Discontinuation of trial treatment for other reasons than PD/death: consider event at time of discontinuation.
\item Supplementary 5: Death $\ge 6$ months after last non-PD response assessment: censored at last available response assessment.
\ei
The assumptions imposed when censoring at an intercurrent event are discussed in Sections~\ref{censoring} and \ref{diffestdiffdis}.

\begin{landscape}
\begin{table}[h]
\begin{center}
{\footnotesize
\begin{tabular}{c|p{2.5cm}|p{1.8cm}|p{7cm}|p{3.5cm}|p{5.5cm}}
  \hline
  Analysis & Population & Variable & Intervention effect & Summary measure & Rationale \\ \hline
  Primary  & FL patients, defined by list of in- and exclusion criteria & Inv-PFS & NALT: {\bf treatment policy} \newline Drop-out, withdrawal: {\bf hypothetical} \newline Death: {\bf composite} & unadjusted hazard ratio and logrank test, stratified by chemotherapy and FLIPI1 & Composite: interest would be in time-to-progression, but we make ``death'' part of a composite. \\ \hline
  ``Primary''  & as primary & IRC-PFS & as primary & as primary &  \\ \hline
  Sensitivity 1 & as primary & as primary & as primary & unadjusted hazard ratio and logrank test, unstratified & \\ \hline
  Sensitivity 2 & as primary & as primary & as primary & unadjusted hazard ratio and logrank test, stratified, using re-randomisation & Assesses sensitivity of stratified log-rank test to dynamic randomisation procedure, see e.g. \cite{kaiser_12}. \\ \hline
  Supplementary 1 & as primary & as primary & Withdrawals prior to PD: {\bf composite} for obinutuzumab, {\bf hypothetical} for rituximab & as primary & Assess impact of loss to follow-up. \\ \hline
  Supplementary 2 & as primary & as primary & Missed assessment prior to PD or clinical cutoff: {\bf composite} & as primary & Assess impact of missed assessments. \\ \hline
  Supplementary 3 & as primary & as primary & NALT prior to PD: {\bf hypothetical} & as primary & Assess potential confounding of the treatment effect estimates by subsequent therapy. See also Section~\ref{diffestdiffdis}. \\ \hline
  Supplementary 4 & as primary & as primary & Discontinuation of trial treatment for other reasons than PD/death: {\bf composite} & as primary & \\ \hline
  Supplementary 5 & as primary & as primary & Death $\ge 6$ months after last non-PD response assessment: {\bf hypothetical} & as primary & \\ \hline
\end{tabular}
}
\caption{List of original sensitivity analyses from GALLIUM protocol. ``IRC-PFS'' stands for PFS as assessed by independent review committee, see the discussion below. The column {\it Analysis} uses the post-addendum terms. Estimand strategies according to the addendum are emphasized in {\bf bold}.}
\label{gallium1}
\end{center}
\end{table}
\end{landscape}

The primary analysis yielded an estimated hazard ratio of 0.66, with $p$-value 0.0012 and 95\% confidence interval from 0.51 to 0.85 \cite{marcus_17}. All but Supplementary Analysis 1 were sufficiently consistent with this result.

Assembling Table~\ref{gallium1} provided the following insights that may be helpful in designing future studies or analysis plans:
\bi
\item It is not entirely obvious how to position IRC-PFS. The primary endpoint of the trial was Inv-PFS, meaning that the timing of the clinical cutoffs for interim analyses was based on the number of events for Inv-PFS, and the $p$-value for this endpoint was primarily considered by the independent data monitoring committee (iDMC) when making their recommendation to either stop or continue the trial at any interim analysis. However, the iDMC charter also asked the iDMC to ascertain that point estimates for Inv- and IRC-PFS were ``consistent''. Furthermore, the protocol stated that {\it Although the primary efficacy endpoint is the investigator-assessed PFS, PFS based on IRC assessments will also be analyzed to support the primary analysis. In the United States, IRC-assessed PFS will be the basis for regulatory decisions.} For that purpose, all the analyses in Table~\ref{gallium1} had been repeated for IRC-PFS and, given their importance for regulatory purposes and since we only alter one aspect of the primary estimand (assessment method for PD), we consider these still sensitivity analyses for Inv-PFS. However, one could argue that these should rather be deemed {\it supplementary} in post-addendum language, as they could be considered targeting an alternative estimand.
\item Each sensitivity analysis only modifies one aspect of the primary analysis at a time, in line with the recommendation in the addendum.
\item For Supplementary Analysis 1, a different strategy for the intercurrent event is used depending on which arm the patient who withdrew was randomized to, so this can be considered a ``worst case'' approach. In hindsight, the scientific objective, the implied estimand, and the estimated treatment effect based on it is thus difficult to interpret, to say the least.
\item For the hypothetical estimand in Supplementary Analysis 3, the analysis strategy specifies censoring at initiation of NALT, as in Section~\ref{diffestdiffdis}. We thus estimate PFS assuming that those patients who received NALT are comparable to those that did not need NALT. As discussed in Section~\ref{censoring}, if the corresponding effect estimate should be unbiased we need to make e.g. the assumption of random censoring, but based on the discussion by Fleming et al. \cite{fleming_09}, this assumption is unlikely to hold. As discussed in Section~\ref{diffestdiffdis}, IPCW or RPSFT would be (complex) options to estimate a hypothetical estimand.
\item In Section~\ref{summary} we discuss that one could argue to split the summary measure in two separate attributes, one for the hypothesis test and one for the effect estimate. In this context it is noteworthy that Sensitivity Analysis 2 indeed only varies the {\it hypothesis test}, but not the {\it effect estimate}.
\item Finally, the choice of the summary measures implies that PH was assumed for all the estimands in Table~\ref{gallium1}, and also Table~\ref{gallium2}. For the primary endpoint, Inv-PFS, this assumption was based on results of the predecessor trial PRIMA \cite{salles_10}. The treatment arm from PRIMA became part of the control arm of GALLIUM, and estimates of the survival function of the treatment arm in PRIMA revealed a remarkably constant hazard of an Inv-PFS event over time, a feature regularly observed in FL. The trial team thus assumed the same shape for the hazard function in the experimental arm in GALLIUM, leading to the PH assumption. And in fact, the analysis of GALLIUM indeed showed quite constant estimated hazard functions and thus the PH assumption seemed justified also in hindsight. Interestingly, the average regression effect as proposed by Xu and O'Quigley \cite{xu_00} was estimated to be 0.68 (computed using the \proglang{R} \cite{R} package \pkg{coxphw} \cite{coxphw}), so quite close to the estimated primary analysis hazard ratio.

    Strictly speaking, the above rationale justifying the PH assumption based on an earlier trial exclusively applies to Inv-PFS in GALLIUM. Similar considerations for all the other analyses in Table~\ref{gallium1} and \ref{gallium2} had not been made at the time.
\ei
The GALLIUM trial also reported a set of results on secondary endpoints \cite{marcus_17}. These are listed in Table~\ref{gallium2} and, in post-addendum language, these would be ``supplementary''. It is remarkable that all these endpoints, i.e. OS, Time to NALT (TTNALT), EFS, DFS, and duration of response (DOR) are basically received by varying handling of clinical events for the variable and intervention effect, as well as the population attribute for DFS and DOR. The analysis strategies for the estimands in Table~\ref{gallium2} are the same as for Table~\ref{gallium1} and in addition:
\bi
\item Clinical cutoff for OS and TTNALT: administratively censor at date last known alive.
\item Supplementary 7: The protocol specified that after PD, patients were to be followed up for NALT and death, allowing to use a treatment policy strategy for PD for TTNALT. 
\ei

\begin{landscape}
\begin{table}[h]
\begin{center}
{\footnotesize
\begin{tabular}{p{3cm}|p{2.5cm}|p{2.5cm}|p{5.5cm}|p{3.5cm}|p{5.5cm}}
  \hline
  Analysis & Population & Variable & Intervention effect & Summary measure & Rationale \\ \hline
  Primary  & FL patients, defined by list of in- and exclusion criteria & Inv-PFS & NALT: {\bf treatment policy} \newline Drop-out, withdrawal: {\bf hypothetical} \newline Death: {\bf composite} & unadjusted hazard ratio and logrank test, stratified by chemotherapy and FLIPI1 & Composite: interest would be in time-to-progression, but we make ``death'' part of a composite. \\ \hline
  Supplementary 6 \newline Overall survival & as primary & OS & PD, NALT: {\bf treatment policy} \newline Drop-out, withdrawal: {\bf hypothetical} & as primary & \\ \hline
  Supplementary 7 \newline Time to NALT & as primary & TTNALT, time from randomisation to death or NALT & PD: {\bf treatment policy} \newline Drop-out, withdrawal: {\bf hypothetical} \newline NALT, death: made part of a {\bf composite} & as primary & \\ \hline
  Supplementary 8 \newline Event-free survival & as primary & EFS, time from randomisation to Inv-PFS event or NALT & Drop-out, withdrawal: {\bf hypothetical} \newline NALT, death: made part of a {\bf composite}  & as primary & \\ \hline
  Supplementary 9 \newline Disease-free survival & Patients with CR prior to NALT & DFS, time from first occurrence of CR to Inv-PFS event & as primary & unadjusted hazard ratio, stratified by chemotherapy and FLIPI1 & Non-randomized comparison between arms, no hypothesis test performed. \\ \hline
  Supplementary 10 \newline Duration of response & Patients with PR or CR prior to NALT & DOR, time from first occurrence of PR or CR to Inv-PFS event & as primary & unadjusted hazard ratio, stratified by chemotherapy and FLIPI1 & Non-randomized comparison between arms, no hypothesis test performed. \\ \hline
\end{tabular}
}
\caption{List of supplementary analyses from GALLIUM protocol. ``PR'' stands for ``partial response''. The column {\it Analysis} uses the post-addendum terms. Estimand strategies according to the addendum are emphasized in {\bf bold}.}
\label{gallium2}
\end{center}
\end{table}
\end{landscape}

\section{Discussion}
\label{discussion}

In this paper, we have outlined an estimand framework for a \te endpoint and discussed all four estimand attributes defined in the addendum as they apply to a \te endpoint. We illustrate that these attributes cannot be seen independently, but have to be considered inter-related. Intercurrent events often make measurements of a \te endpoint incomplete, or even impossible to observe in the case of an intercurrent event of death. Depending on the targeted estimand, various approaches can be used to estimate it. For a hypothetical estimand, simple censoring at the intercurrent event may often yield a potential estimator, but at the price of the strong assumption that the endpoint for those patients experiencing the intercurrent event can be ``imputed'' using data from patients with longer follow-up, but not experiencing the intercurrent event. Alternatively, methods developed in the causal inference literature can be used to estimate a hypothetical estimand, making alternative assumptions and necessitating data collection on factors that are prognostic and predict the intercurrent event, e.g. a treatment switch \cite{dodd_17}.

As summary measures, the logrank test and the hazard ratio based on the CPHR are overwhelmingly used today in the routine analysis of trials with \te endpoints. However, both these methods are intimately connected to the PH assumption. While the PH assumption may often be made based on previous or external knowledge, as e.g. in GALLIUM in Section~\ref{gallium}, there are indications where it is clear that it generally does not apply, e.g. immuno-oncology \cite{chen_13}. Still, trials are often powered based on the logrank test and the pre-specified effect quantifier is based on CPHR. This asks for a choice between the two potential ways of moving forward in the context of the addendum as outlined in Section~\ref{intervention}.
\bi
\item Making testing and estimation consistent even under NPH would methodologically be possible, using e.g. one of the estimators discussed in Section~\ref{coxproblem}. The tradeoff would be a major logistical and educational effort for all parties involved, i.e. rewriting of programming and reporting templates and education of statisticians, clinicians, HAs, and the broader scientific community on the application and interpretation of these methods. Also, if, as an example, the average regression effect approach by Xu and O'Quigley \cite{xu_00} would be adopted, methodology would first have to be developed e.g. for sample size computation, sequential monitoring of a trial, estimation of this average effect from interval censored data, etc.
\item The other option to move forward would be the gatekeeper approach, which is at least informally accepted by HAs today.
\ei If this second approach remains valid moving forward, it would be crucial for trial sponsors to understand:
\bi
\item Even if the hypothesis test can be different from the effect quantifier - is validity of the chosen hypothesis test enough to justify its application, or does the hypothesis test still need to refer to a clearly defined estimand? This would make the use of the logrank test questionable under NPH.
\item What are the criteria that will make HAs accept an effect quantifier? As suggested by a reviewer, could it be an option to base the hypothesis test on an analysis method that makes a set of plausible but minimal assumptions, but then use one or more effect quantifiers in the package insert making different and/or stronger assumptions, e.g. proportional hazards or even parametric?
\ei
A discussion of these aspects, specifically the role and interplay of testing and estimation, in the ICH E9 final addendum or general guidance by HAs on these aspects would certainly help sponsors to set up RCTs with \te endpoints in the future, especially when NPH is anticipated.

Irrespective of which approach is favoured, a general comment on the causal interpretation of \te data and hazard ratios is in order. As discussed by Akacha \cite{akacha_17rss}, while the addendum is not explicitly mentioning the word ``causal'', reference to ``causal thinking'' is made implicitly via referencing {\it potential outcomes} (Section~A.3.1) and adoption of the principal stratum strategy. However, as e.g. Hernan \cite{hernan_10} and Aalen et al. \cite{aalen_15} point out, the use of the hazard ratio for causal inference is not straightforward, even in the ideal situation of PH and absence of unmeasured confounding and measurement error. Furthermore, as discussed by Hernan (\cite{hernan_17}, Fine point 17.1), truncation of a \te variable by competing events raises logical questions about the meaning of a causal estimand, and these issues cannot be bypassed by statistical techniques. So, even when perfectly adopting the addendum framework for \te endpoints in the future and PH being fulfilled, validity of causal conclusions for a \te endpoint might remain unclear, even in an RCT, and needs further research. We would also welcome if the addendum in its final form could be more clear on whether estimands have to be constructed in a way that they allow for causal statements, or alternatively, if the addendum can be interpreted more in the sense of an ``operational guidance'' of how to set up RCTs well and transparently.

Interestingly, as discussed by Aalen et al. \cite{aalen_15}, accelerated failure time models could potentially provide both, effect estimates that are robust against the censoring distribution and allow for a causal interpretation.

We have discussed the current heterogeneity in variable definitions and the DATECAN initiative that aims at unifying these definitions in many indications in oncology. On top of that, harmonization efforts are underway for response definitions, e.g. extending the {\it Response Evaluation Criteria In Solid Tumours} (RECIST) guideline to testing drug agents in immunotherapy \cite{seymour_17}. We are definitively supportive of all these efforts and hope that these will be extended to other disease areas as well. However, with a precisely defined endpoint and strategy how to handle intercurrent events, only part of what constitutes an estimand according to the addendum is specified. We believe that the estimand framework will further help to align trial objectives and quantification and interpretation of effects. In addition, having the estimand discussion within teams at trial onset will help to define the data collection strategy. If a hypothetical estimand is targeted e.g. for OS, data collection strategy becomes especially important: data on prognostic factors for experiencing an intercurrent event, e.g initiation of NALT or a switch from control to treatment arm, needs to collected. Hudis et al. provide an illustrative discussion on these aspects in the penultimate paragraph of their paper \cite{hudis_07}. One aspect is important to note in this context: for potentially lethal diseases like many cancers, patients remain under treatment also after progression, often even within the initial trial protocol, so that follow-up for OS is feasible. Such comprehensive follow-up for OS might be more difficult to achieve in other indications, e.g. Alzheimer's disease. Experience shows that if patients stop treatment, due to the high burden a visit to the hospital with all the trial assessments is for the patient and caretakers, the odds that patients completely drop-out of the trial after stopping treatment are higher than in many oncology indications.

We illustrate our points with two case studies from oncology, and we show that familiar endpoints such as PFS, TTP, or EFS can be simply interpreted as different estimands.

Strict compliance to the estimand methodology outlined in the addendum is desirable moving forward. However, as remarked by a reviewer, this may increase costs, e.g. because of power loss when using the treatment policy strategy for an intercurrent event or diluting endpoints when using the composite strategy.

Regulatory guidelines, e.g. the {\it FDA Guidance for Industry: Clinical Trail Endpoints the Approval of Cancer Drugs and Biologics} \cite{fda_endpoints}, specify definitions of endpoints such as PFS or TTP, and also outline potential sensitivity analyses. This guidance is reflected in the list of analysis for GALLIUM in Tables~\ref{gallium1} and \ref{gallium2}. However, the guidance currently does neither discuss the assumptions that are (implicitly) made on censoring nor are estimands specified from which the proposed analyses, or estimators, are derived. We anticipate that regulatory guidelines will have to updated to reflect the changes brought to ICH E9 by the addendum.

Based on our experience and working on the tables in Section~\ref{gallium}, and similar to the recommendations by Hernan and Robins \cite{hernan_17nejm} in the context of pragmatic trials, we recommend that when defining an estimand for a trial with a \te endpoint, the sponsor should
\bi
\item involve all relevant stakeholders of the trial, i.e. statisticians, clinicians, internal regulatory colleagues, health authorities, payer organizations if the trial is industry-sponsored, and maybe even patients,
\item identify all endpoint-defining and potential intercurrent events,
\item define for each intercurrent event the estimand strategy to be applied, ideally by inferring this strategy from a precisely formulated trial objective,
\item tabulate all these events and determine for each of them an estimand strategy,
\item and finally apply or develop methodology that can estimate the chosen estimand, with a precise definition of the estimator and a description and discussion of the assumptions made, e.g. with respect to censoring.
\ei
By doing so, the sponsor should make sure to take into account aspects specific to the disease under study and make an effort to match the definition with either guidelines in the field, such as DATECAN, CONSORT, or RECIST, or other clinical trials. This will facilitate discussions on regulatory approval, allow for easier comparison of trials now and in the future, and allow the use of trial data as historical controls, in meta-analyses, or surrogacy assessments.

\section{Acknowledgments}
\label{ack}
I would like to thank the reviewers for their generous and constructive comments that substantially improved this paper. Specifically, I would like to thank one reviewer for pointing out the lack of a clear estimand implied by CPHR. This feedback made me extend Section~\ref{summary} in the revision.

I would like to acknowledge numerous discussions on the topic with Mouna Akacha, Hans Ulrich Burger, James Roger, and Marcel Wolbers and thank Chris Harbron and Tina Nielson for proofreading an earlier version of this paper.

\bibliographystyle{ieeetr}
\bibliography{stat}

\end{document}